\documentclass[runningheads]{llncs}
\usepackage{graphicx}
\usepackage{biblatex} 
\begin{document}
%


\title{Motivational models for validating agile requirements in Software Engineering subjects}
%
%
\author{Eduardo A. Oliveira\inst{1} \and Leon Sterling\inst{1,2}}
\authorrunning{Oliveira, E.A. and Sterling, L.}
%
\institute{University of Melbourne, Parkville 3010, VIC, AU \\ 
Swinburne University of Technology, Hawthorn, 3022, VIC, AU\\ 
\email{eduardo.oliveira,leonss@unimelb.edu.au}}

\maketitle              
\begin{abstract}
This paper describes how motivational models can be used to cross check agile requirements artifacts to improve consistency and completeness of software requirements. Motivational models provide a high level understanding of the purposes of a software system. They complement personas and user stories which focus more on user needs rather than on system features. We present an exploratory case study sought to understand how software engineering students could use motivational models to create better requirements artifacts so they are understandable to non-technical users, easily understood by developers, and are consistent with each other. Nine consistency principles were created as an outcome of our study and are now successfully adopted by software engineering students at the University of Melbourne to ensure consistency between motivational models, personas, and user stories in requirements engineering. 

\end{abstract}
\section{Introduction}
\label{sec:introduction}
\noindent Software development practices have changed over the last twenty years, with agile development more prevalent than traditional waterfall development. While the agile manifesto prefers working code over documentation, there is a need for models that can help communicate and document a high level understanding. Clear models are particularly important as software development teams are increasingly multi-disciplinary due to the expansion of application domains where software is essential. The increase in disciplines places greater demands on communication as less knowledge can be assumed. Both technical and non-technical team members are able -- and often expected -- to contribute to ongoing discussions and decisions about requirements. 

In this paper, we present a case study on how Motivational Modelling (MM) can be used effectively to support the creation of better agile requirements artifacts by software engineering students at the University of Melbourne. As an outcome of our analysis in the presented case study, nine Consistency Principles (CP) were created to support the cross validation between motivational models, personas, and user stories in Requirements Engineering (RE).

\section{Background}
\noindent RE practices in agile software development involve extensive collaboration and face-to-face communication \cite{inayat2015systematic}. The Agile Manifesto states that priority should be given to ‘‘\textit{individuals and interaction over processes and tools, working software over comprehensive documentation, customer collaboration over contract negotiation, and responding to changes over following a plan}’’ \cite{beck2001manifesto}. Requirements are the base of all software products and consequently RE plays an important role in system development.

Overall high-level project scope is discussed and defined upfront with the customer and is revisited in each iteration (e.g. using Scrum). Requirements are discussed, prioritised, planned, monitored and reviewed until the end of the iteration (or sprint). The cycle is repeated every iteration until the product is finalised. It is essential that customers and developers work together during the whole software engineering process to achieve a clear understanding of requirements and validate the product under development.

Previous studies have identified several challenges related to RE in the past few years as this is one of the most critical phases for project success \cite{schon2017agile,inayat2015systematic,bhat2006overcoming}. Face-to-face communication, customer involvement and interaction are challenges discussed among researchers as RE often involves different stakeholders with diverse backgrounds.

Personas and user stories can be used to specify customer requirements \cite{cohn2004user}. A persona is a description of an archetypical user of a software system, created from research about the potential real users of the software system. A user story description consists of a statement, among other things. To be amenable to humans as well as to machines, a user story statement should relate to both a persona and a goal \cite{kamthan2015using}. A goal is an intended outcome of a persona interacting with a software system. Personas provide the rationale for the existence of user stories. The success of software products is highly dependent on validations performed on users' goals. The relationships between personas, user stories and goals are critical. Non-technical approaches can be used to facilitate communication and better overall understanding among stakeholders when combined appropriately \cite{daneva2013agile,blomkvist2015communication}. 

In recent years we have used MM to effectively bridge requirements artifacts such as system goals, personas and user stories consistently. They improve communication between students and industry partners. Recently we have been using MM to validate requirements artifacts during the requirements elicitation and design phases of software engineering subjects at the University of Melbourne. 

\section{Motivational Modelling}
\label{sec:mm}

\noindent Motivational modelling is a method to understand a problem or desired system that is accessible to all stakeholders \cite{sterling2009art}. It was originally viewed as an extension of traditional RE methods to agent-oriented systems by placing more emphasis on goals and roles. The modelling methods subsequently evolved to encompass a broader range of quality goals, and to explicitly include emotional factors, see for example  \cite{Sterling2018EmergencyAlarms,quitch}. 

Motivational models present a hierarchical diagram of the goals of a system at a high level of abstraction. The modelling notation uses four main elements: (i) roles which represent the stakeholders, (ii) functional requirements which list what the system must achieve, (iii) quality (or non-functional) requirements to list issues such as performance, security and scalability, and (iv) emotional requirements which list both positive feelings people want to have when interacting with the system, and possible concerns. We use the terms \textit{do goals}, \textit{be goals} and \textit{feel goals} to express functional, quality, and emotional requirements, with the words `goals' and `requirements' often used interchangeably. More informal terminology helps software developers engage with non-technical clients and stakeholders.

A good visual notation is helpful for the models to be understood and welcomed by users. We advocate the icons that are presented in Figure~\ref{fig:team8}. Parallelograms are used to represent what the system should do; clouds are used to represent how the system should be, stick figures represent the roles/stakeholders in the system, while hearts, naturally enough, represent the emotions that are trying to be engendered in the stakeholders.

Motivational models can be developed systematically in a two-stage process, the method for which is described in \cite{Lopez-Lorca}. Since the focus in the paper is on using a motivational model for validation, we present a case study in how to check the consistency of personas and user stories with motivational models using our proposed nine consistency principles.

We have been teaching software engineering subjects for many years. A feature is the use of real clients for projects. A major challenge for students is to elicit requirements from (often non-technical) clients to get a clear description of requirements that is understood by both student developers and clients.
In this paper, we present and discuss an exploratory case study sought to understand how software engineering students could improve the quality of their requirements through the use of MM. The research question investigated in this study was: How to effectively support the creation of better agile requirements artifacts so they are understandable to non-technical users, easily understood by developers, and are consistent with each other?
We expected to identify advice to students in the form of consistency principles for requirements artifacts. 

\section{Method}
\subsection{Context}
Motivational modelling has been taught to software engineering students within three subjects which are part of the Master of Engineering (Software) since 2017 and, more recently, to an IT project subject within the Masters of Information Technology at the University of Melbourne. A MM tool was developed by software engineering students from the same university in 2018 to facilitate creation and design of motivational models that are used during the requirements elaboration phase of software development. The tool has been continuously updated by new students in ongoing research. Our software engineering students develop motivational models for their real-world projects during the requirements elaboration phase, which we refer to as the Inception sprint.

\begin{figure}[!t]
   \centering
        \includegraphics[width=1\columnwidth]{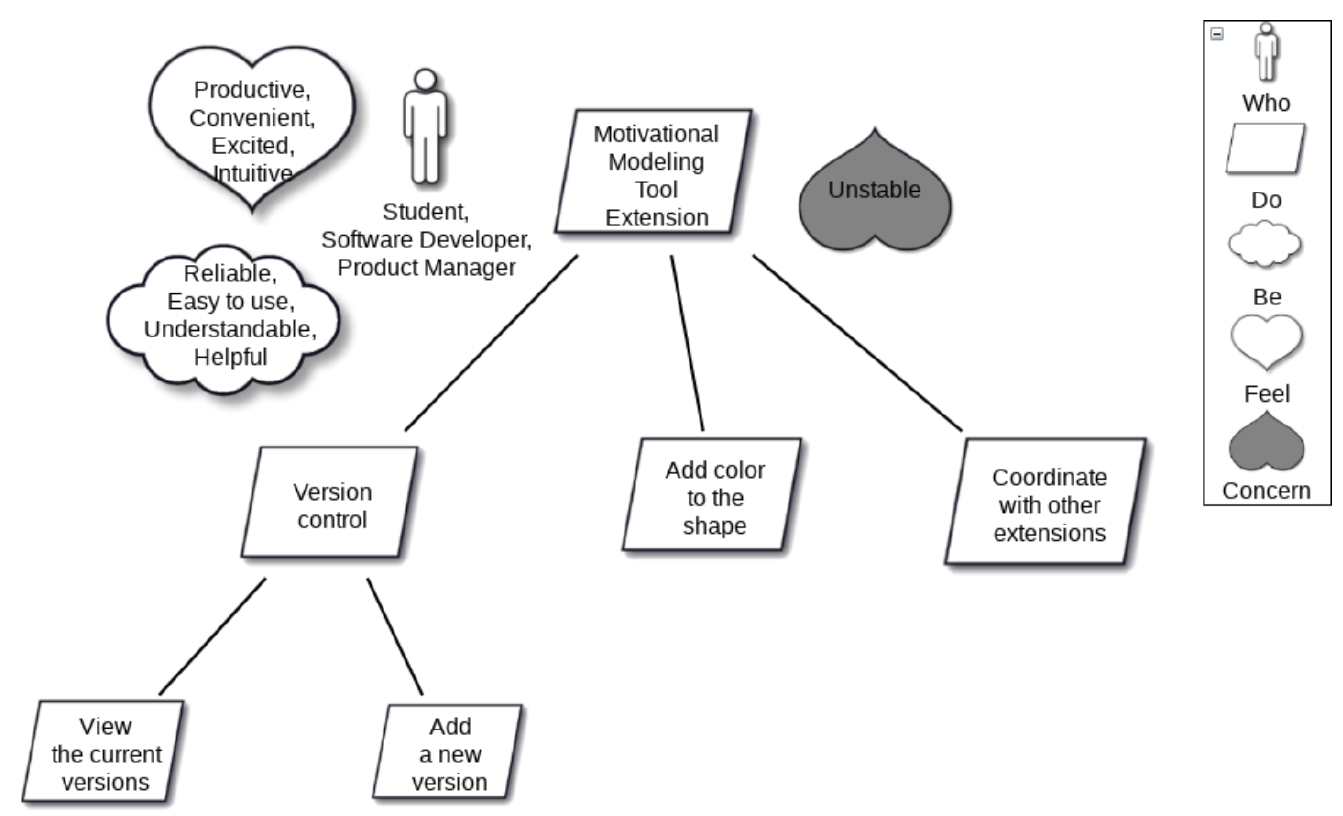}
        \caption{Motivational modelling extension}
        \label{fig:team8}
\end{figure}

\subsection{Creating a Motivational Model}
In the IT project subject, the project goal of the four student team was to extend the existing MM tool with new features that could facilitate communication with industry partners and the quality (completeness and consistency) of personas and user stories produced during Inception sprint.

The motivation model for the extensions developed by the students is given in Figure~\ref{fig:team8}. In essence, the model is a problem description rather than a requirements specification.

The six parallelograms give a high level view of the functional requirements identified as possible extensions to the existing MM tool. The overall goal of the project is given in the root of the tree at the top of the page in Figure~\ref{fig:team8}, namely to create an extension to the MM tool (currently hosted at https://qs.motivationalmodelling.com/). There are three components of the extension, expressed in the three subgoals of the main overall goal. The first component is to provide version control, which in turn has two components, \textit{View the current versions}, and \textit{Add a new version}. The second component is \textit{Add color to the shape}. The third, and more minor, component is to \textit{Coordinate with other extensions}, which may entail some more code.

The roles being considered are given in the label under the person figure in Figure~\ref{fig:team8}. The roles are \textit{Student}, \textit{Software Developer}, and \textit{Product Manager}.

There are four terms in the cloud in Figure~\ref{fig:team8}. The terms represent the quality goals of the project, namely that the extension (and the MM tool more generally) should be \textit{Reliable}, \textit{Easy to use}, \textit{Understandable}, and \textit{Helpful}. Because all of the quality goals apply to the whole project, they are listed in a single cloud. Similarly, the four terms in the heart in Figure~\ref{fig:team8} represent the emotional goals of the whole project. We note there is a concern expressed in the upside-down heart that the system may be unstable. However the concern will take no part in the validation process.

Motivational models such as in Figure~\ref{fig:team8} can be considered as annotated graphs. The graph is almost always drawn as a tree with the root of the tree at the top of the page. We refer to \textit{do goals} in the motivational model as nodes in the tree where convenient in the discussion below.  

This case study checked the consistency between the motivational model developed by the students with their personas and user stories. The consistency between artifacts were captured through documents created by students and uploaded to Confluence during Inception sprint.

\section{Results}
As an outcome of our analysis in the presented case study, nine Consistency Principles (CP) were created to support the cross validation between user stories and personas:
\begin{itemize}
\item CP1: There should be at least one persona for every role in the motivational model.  
\item CP2: Every persona should relate to a role in the motivational model.
\item CP3: The grouping of user stories into epics should be reflected in the motivational models.
\item CP4: Every user story needs to relate to one of the nodes in the model.
\item CP5: There should be at least one user story for every leaf node in the motivational model. 
\item CP6: There should be at least one user story for every quality goal.
\item CP7: Every quality goal occurring within a user story should appear in the motivational model.
\item CP8: There should be at least one user story for every emotional goal.
\item CP9: Every emotional goal occurring within a user story should appear in the motivational model.
\end{itemize}

The consistency principles started to be identified during the use of motivational models by over 500 students in several subjects since 2019 once the tool was started to be used more frequently by students. The teaching team identified inconsistencies between the Inception artifacts, namely motivational models, personas and user stories in several offers of software engineering project-based subjects. They also observed noisy/unclear communication in regards to project goals between students and clients. The teaching team redesigned assessment criteria for Inception sprint to support students during the development of RE artifacts. 
The ultimate goal in redesigning assessment criteria for Inception sprint was (i) to improve communication between different stakeholders with diverse backgrounds and, (ii) to stop assessing RE artifacts independently and individually. The Inception sprint main focus changed from requirements specification to problem description and understanding. The use of MM diagrams were highly encouraged as the first task students should get completed during requirements elaboration phase. After agreeing (students and clients) on a high level diagram of the goals of a system, students could then use that same artifact to develop the next RE artifacts sequentially and not independently. MM diagrams became the center piece to connect RE artifacts during Inception.
In the next section, we explain how CP were explicitly used in this case study as an afterwards validation process.

\subsection{Identifying Consistency Principles Through the Case Study Analysis}

There are three roles in the MM developed by participants for their extension project (Figure~\ref{fig:team8}): student, software engineer and product manager. In fact, the team had two personas for student, representing students of different ability and enthusiasm. Whether two separate roles are needed is a question to be discussed with the project client, but it is positive that the question has been raised. So the check is twofold. These considerations of the roles gives rise to two Consistency Principles -- \textbf{CP1} and \textbf{CP2}. 

\begin{figure}[ht]
   \centering
 \includegraphics[width=1\columnwidth]{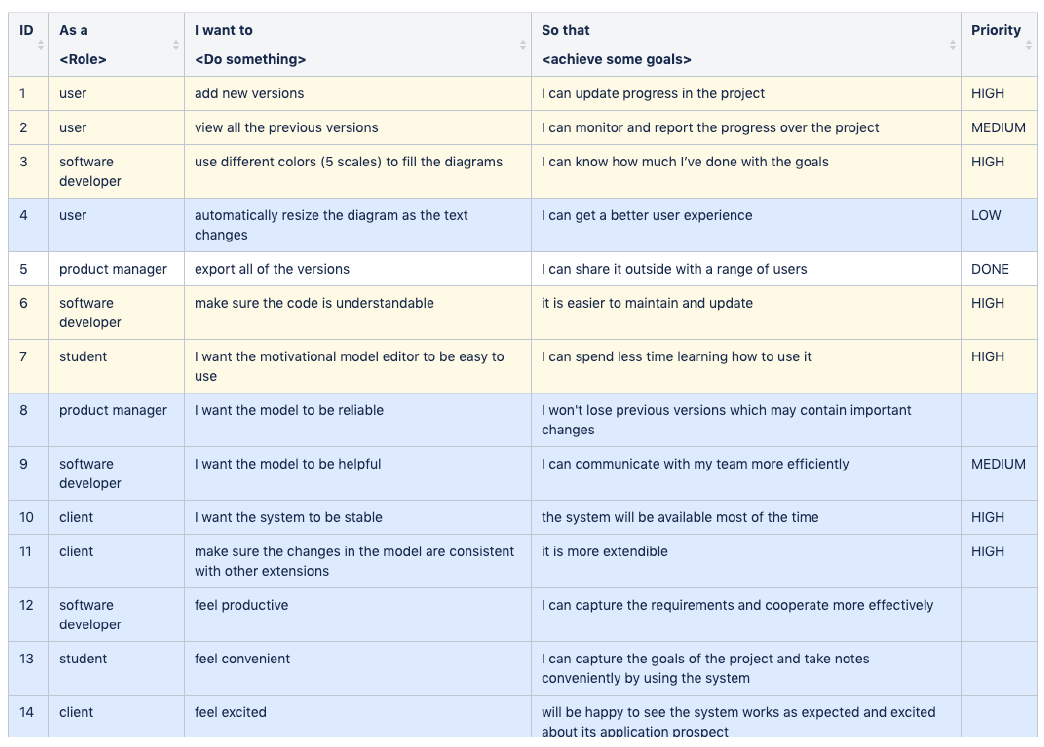}
         \caption{User stories for the Motivational Modelling Editor extension project}
         \label{fig:userstories}
 \end{figure}

The checking of the user stories is more complicated, which we now demonstrate. The student team produced a list of fourteen user stories (Figure~\ref{fig:userstories}). The different row colours are related to the development status of the user stories (to do, doing, done) on JIRA, and are not relevant to this discussion. The list of user stories was accepted by the team supervisor. We demonstrate how the user stories and motivational model can both be improved by doing mutual cross-checking and validation. 
Let us consider the first three user stories from Figure~\ref{fig:userstories}: (i) \textit{As a user, I want to add new versions so that I can update progress in the project}, (ii)\textit{As a user, I want to view all the previous versions so that I can monitor and report the progress over the project} and, (iii)\textit{As a software developer, I want to use different colors (5 scales) to fill the diagrams so that I can know how much I've done with the goals}.

At first glance the stories seem appropriate (even though inconsistent with CP2 as `user' is different from identified roles in MM -- student, software developer and product manager). The three user stories are related to the three leftmost leaf nodes of the MM. In general, that is a good practice and gives rise to \textbf{CP4} and \textbf{CP5}.

The purpose of the user story encapsulated in the `so that' phrase shows up some limitations. If the purpose is to monitor progress, as expressed in the first user story, then surely \textit{monitor progress} should be a goal in the MM, but it is not. Worse in fact, the purpose of colouring the nodes expressed in the third user story is to monitor progress in the project. So that all three user stories should be re-written and the MM re-drawn. 

Re-structuring the user stories effectively restructures the epics that encompass the user stories. Systematically including every goal in the MM can make the MM unduly complicated or unwieldy. In that case the model should be split into a hierarchy of models. A natural division would be a MM for each epic consisting of several user stories. As we move around user stories, we are effectively modifying the epics. The observation leads to \textbf{CP3}.

User story 4 represents a straight omission from the MM. The user story reads \textit{As a user, I want to automatically resize the diagram as the text changes, so that I can get a better user experience}. This user story was added to the project and is an example of coordinating with other features of the MM tool editor. A new node \textit{Automatic resizing of do goals to fit text} should have been added to the \textit{Coordinate} goal. Furthermore, the \textit{Coordinate} goals would be better re-worded as coordinate with other features rather than with other extensions. The MM should have been updated.

User story 6 gives a user story relating to \textit{Understandable}. User story 7 gives a user story relating to \textit{Ease of use}, which are validated through \textbf{CP6} and \textbf{CP7}.

Quality goals and emotional goals are handled similarly in user stories. The two principles for quality goals should be adapted for emotional goals (\textbf{CP8} and \textbf{CP9}). 

After identifying and consolidating the nine Consistency Principles, we have incorporated them in assessment rubrics for teaching in 2021.

\section{Conclusions}
\label{sec:conclusion}

\noindent We have described nine consistency principles which can be used to cross check motivational models with user stories and personas to ensure consistency and completeness of user requirements. The models can also help to hierarchically describe a project and develop a data dictionary, the latter feature not discussed in the paper. The methods have been used successfully at the University of Melbourne for several years. 

The MM tool is currently being extended to allow students to extract personas and user stories from MMs, to add colours and notes to the diagram. Users will also be able to upload personas, user stories and epics (text) to the tool, which will convert them into a MM diagram. Future studies should focus on further examining the impact of the consistency principles on requirements quality and on communication between stakeholders.

\section*{Acknowledgments}
The authors would like to thank the teaching staff and students doing the Software Engineering subjects at the University of Melbourne who have contributed to the ideas of the paper. The work was partially supported by ARC Discovery grant DP200102955, `Maturing design-led innovation processes with motivational models'.
%
%
%
\printbibliography[title={References}]
\end{document}